\documentclass[
    aps,
    pra,
    reprint,
    groupedaddress,
    twocolumn
]{revtex4-2}

\usepackage[T1]{fontenc}

\usepackage{amsmath}
\usepackage{amssymb}
\usepackage{mathtools}
\usepackage{braket}
\usepackage{dsfont}
\usepackage{booktabs}

\usepackage{graphicx}
\graphicspath{{./images/}}
\usepackage{xcolor}

\usepackage{natbib}

\usepackage{algpseudocode}

\newcounter{algorithm}
\renewcommand{\thealgorithm}{\arabic{algorithm}}

\newcommand{\algorithmcaption}[1]{%
    \refstepcounter{algorithm}%
    \noindent\textbf{Algorithm \thealgorithm.} #1\par\medskip
}

\usepackage{flushend}
\usepackage{placeins}

\begin{document}

\title{Heuristic Lookahead Distillation Protocol Search}

\author{Matthew Barber}
\author{Stefano Pirandola}
\affiliation{Department of Computer Science, University of York, York YO10 5GH, United Kingdom}

\begin{abstract}
    Bipartite qubit entanglement distillation is the process of converting noisy ebits into pure ebits using only local operations and classical communication.
    This is a core operation for quantum repeaters, enabling such crucial tasks as long-distance quantum communication and distributed quantum computing.
    In this work, we introduce a method for searching for entanglement distillation protocols and, using this technique, distil qubit Werner states at a higher rate than could be achieved using previously discovered protocols.
    In particular, we demonstrate the advantage of our new distillation strategy by improving the best known lower bound for the two-way-assisted quantum capacity of the qubit depolarising channel across a wide range of channel parameters, making progress in one of the long-standing problems of quantum information theory.
\end{abstract}

\maketitle

\section{Introduction}

Entanglement is one of the key resources of quantum communication and is required for many of the most important quantum communication protocols \cite{QuantumTeleportation, SuperDenseCoding, BB84, Ekert1991, pirandola2020advances}.
However, entanglement can only be created locally \cite{EntanglementReview}.
Therefore, for two separated parties to come to share bi-partite entanglement, the two halves of an entangled pair must be distributed to the two parties.
As the pairs are distributed, they will interact with their environments, introducing noise into the system that corrupts the entanglement. Indeed, the distribution of entanglement is affected by fundamental rate-loss limitations both in repeaterless~\cite{pirandola2017fundamental} and repeater-based~\cite{Pirandola2019} scenarios.
\par
The basic unit of entanglement is the entangled bit, or ebit, which is a maximally entangled qubit pair~\cite{Simeone2026CQIT}.
Let us suppose that we have many noisy ebits – that is, qubit pairs whose entanglement is imperfect due to noise in the system.
Although we cannot produce new entanglement non-locally, we can hope to distil the entanglement present into a smaller number of perfect ebits, or states whose fidelity to an ebit converges to $1$ as the number of noisy ebits available to distil goes to infinity.
Naturally, we would like to distil as many true ebits as possible from as few noisy ebits as possible.
Therefore, we measure the efficiency of an entanglement distillation protocol on a particular qubit pair state by the yield of the protocol on that state, defined as the average number of ebits distilled per noisy ebit consumed.
\par
A basic entanglement distillation protocol was described in Ref.~\cite{BBPSSW}. The yield of this protocol is equal to the coherent information~\cite{PhysRevA.55.1613,DevetakShor2005,PhysRevLett.102.210501} of the qubit pairs it is applied to, when the coherent information is positive.
Since then, improved protocols have been developed.
To our knowledge, the best known bipartite entanglement distillation protocols are those of Refs.~\cite{AJOCSS} and~\cite{HDDM}.
\par
In Ref.~\cite{HDDM}, the authors not only introduce a new distillation protocol but describe a framework in which many distillation protocols can be described, including those of Refs.~\cite{AJOCSS,BBPSSW,DEJMPS,HDDM,VV}.
In this framework, we begin with many independent qubit pairs in identical Bell-diagonal states.
We can therefore think of the collection of pairs as being an unknown sequence of Bell states.
We can distil the states by identifying this sequence.
The authors describe two techniques, known as appended ebit measurements (AEMs) and bilateral Pauli measurements (BPMs), that can be used to gain information about this sequence at the cost of some entanglement.
\par
Our main contribution in this paper is to describe a new way of selecting such AEMs and BPMs to identify the sequence of Bell states at a low entanglement cost.
In short, we search the space of potential AEMs and BPMs much like a chess engine does the space of potential chess moves up to a certain depth.
At each step in the search we update the set of `legal moves' in the form of AEMs and BPMs and, once we reach our maximum search depth, we apply a heuristic to evaluate how favourable that particular sequence of measurements was.
\par
In Sec.~\ref{Twirling}, we will describe the well-known technique of twirling that allows us to convert any qubit pair state into a Bell-diagonal state.
In Sec.~\ref{ParityChecks}, we will give a brief description of the framework of Ref.~\cite{HDDM}.
Then, in Secs.~\ref{UpdatingShortlists} and~\ref{RecedingHorizonSearch}, we will describe our new technique for finding efficient distillation protocols in the framework of Ref.~\cite{HDDM}.
Finally, in Sec.~\ref{Applications}, we will show how this technique can be used to achieve better performance than that achieved by the protocols of Refs.~\cite{AJOCSS,HDDM} in distilling qubit Werner states~\cite{WernerState}.
In this way, we improve the best known lower bound for the two-way-assisted quantum capacity or entanglement distribution capacity of the qubit depolarising channel, a long-standing problem in quantum information theory to which only partial solutions have yet been found~\cite{pirandola2017fundamental}.
Table~\ref{tab:main-notation} summarizes our notation.

\begin{table*}
\centering
\caption{Notation used in our paper.}
\label{tab:main-notation}
\footnotesize
\setlength{\tabcolsep}{4pt}
\renewcommand{\arraystretch}{0.98}
\begin{tabular}{@{}p{0.25\textwidth} p{0.70\textwidth}@{}}
\toprule
Symbol & Description \\
\midrule
\multicolumn{2}{@{}l}{\textbf{States}} \\
$\lvert\Phi_{ij}\rangle$ & Bell basis state labeled by bits $i,j$ \\
$\rho$ & Initial density operator of one qubit pair \\
$\rho_{ij}$ & Bell-basis probability of $\lvert\Phi_{ij}\rangle$ in $\rho$ \\
$\rho^n$ & Joint state of the $n$ initial pairs \\
$\sigma$ & Current distribution over Bell-sequence equivalence classes \\
$\sigma_x$ & Random parity of the current state for check $x$ \\
$X_{\sigma,a,i}$, $Y_{\sigma,b,i}$ & Post-check distributions for outcome $i$ of an AEM or BPM \\
\addlinespace
\multicolumn{2}{@{}l}{\textbf{Parameters}} \\
$n$ & Number of initial independent qubit pairs \\
$m$ & Number of qubit pairs remaining in the current system \\
$r$ & Number of copies joined when one pair remains \\
$d$ & Maximum lookahead search depth \\
\addlinespace
\multicolumn{2}{@{}l}{\textbf{Parity checks}} \\
$x$, $y$ & Check vector and Bell-state label sequence, respectively \\
$P$ & Binary matrix defining the symplectic form \\
AEM & Appended ebit measurement \\
BPM & Bilateral Pauli measurement \\
$A$, $B$ & Subspaces of shortlisted AEM and BPM checks \\
$\widetilde A$, $\widetilde B$ & Ordered bases for the AEM and BPM shortlists \\
$m_A$, $m_B$ & Dimensions of the AEM and BPM shortlist spaces \\
$a$, $b$ & AEM and BPM check vectors in $A$ and $B$ \\
$c$ & Performed or candidate check vector \\
$\widetilde K_l$ & Standard basis for the initial shortlists of $l$ pairs \\
\addlinespace
\multicolumn{2}{@{}l}{\textbf{Functions}} \\
$f$ & Removes redundancy from a shortlist basis \\
$g$ & Restricts a shortlist basis to checks commuting with a check \\
$H$ & Base-2 Shannon entropy \\
$H^\star$ & Bounded-entropy estimate of the remaining check count \\
$C$ & Expected remaining parity check count for the protocol \\
$\widetilde C$ & Lookahead estimate of the remaining parity check count \\
$D$, $E$ & Estimated check counts starting with an AEM or BPM \\
\bottomrule
\end{tabular}
\end{table*}

\section{Pauli-Based Twirling}\label{Twirling}

Suppose two parties, Alice and Bob, share a qubit pair.
Let $\Ket{0_{A}}$ and $\Ket{1_{A}}$ denote the two computational basis states for Alice's qubit and $\Ket{0_{B}}$ and $\Ket{1_{B}}$ denote the same for Bob's qubit.
Now, let
\begin{equation}
    \Ket{\Phi_{00}} = \frac{\Ket{0_{A}}\Ket{0_{B}}+\Ket{1_{A}}\Ket{1_{B}}}{\sqrt{2}}\text{,}
\end{equation}
\begin{equation}
    \Ket{\Phi_{01}} = \frac{\Ket{0_{A}}\Ket{1_{B}}+\Ket{1_{A}}\Ket{0_{B}}}{\sqrt{2}}
\end{equation}
\begin{equation}
    \Ket{\Phi_{10}} = \frac{\Ket{0_{A}}\Ket{0_{B}}-\Ket{1_{A}}\Ket{1_{B}}}{\sqrt{2}}\text{,}
\end{equation}
and
\begin{equation}
    \Ket{\Phi_{11}} = \frac{\Ket{0_{A}}\Ket{1_{B}}-\Ket{1_{A}}\Ket{0_{B}}}{\sqrt{2}}\text{.}
\end{equation}
These four Bell states then form a basis for the Hilbert space of the qubit pair.
\par
Now, suppose that Alice and Bob are trying to distil independent and identical qubit pairs, whose density operator we will call $\rho$.
Then, for every $\left(i, j, k, l\right) \in \left\{0, 1\right\}^{4}$, let $\rho_{ijkl}$ be the components of $\rho$ in our basis of Bell states so that
\begin{equation}
    \rho = \sum_{\left(i, j, k , l\right)\in\left\{0, 1\right\}^{4}}{\rho_{ijkl}\Ket{\Phi_{ij}}\Bra{\Phi_{kl}}}\text{.}
\end{equation}
Suppose that our two parties randomly choose between the identity operation, the bit flip operation, the phase flip operation and the bit-phase flip operation with equal probability, agreeing on this choice using classical communication.
Then, they both apply the chosen operation to their respective qubits.
If they forget their choice, the resulting density operator is
\begin{equation}
    \tilde{\rho} = \sum_{\left(i, j\right)\in \left\{0, 1\right\}^{2}} {\rho_{ijij}\Ket{\Phi_{ij}}\Bra{\Phi_{ij}}}\text{.}
\end{equation}
\par
In other words, by performing a random bilateral Pauli operation, Alice and Bob can convert an arbitrary qubit pair into a Bell-diagonal state, keeping the diagonal elements in the Bell basis the same, using only local operations and classical communication (LOCC).
This twirling operation converts our qubit pairs into a useful standard form at the beginning of our distillation protocols.
Without loss of generality, then, in what follows we will assume that $\rho = \tilde{\rho}$ and, for $\left(i, j\right) \in \left\{0, 1\right\}^{2}$, let $\rho_{ij} = \rho_{ijij}$.

\section{Parity Checks}\label{ParityChecks}

Suppose Alice and Bob share $n$ independent copies of $\rho$, for some $n \in \mathbb{N}$.
Then, their joint state is
\begin{equation}
    \rho^{n} = \sum_{\left(i_{1}, j_{1}, \dots, i_{n}, j_{n}\right) \in \left\{0, 1\right\}^{2n}}{\bigotimes_{k=1}^{n}{\rho_{i_{k}j_{k}}\Ket{\Phi_{i_{k}j_{k}}}\Bra{\Phi_{i_{k}j_{k}}}}}\text{,}
\end{equation}
where $\bigotimes$ denotes the tensor product.
That is, their joint state is a classical mixture of sequences of $n$ Bell states.
If they can identify the sequence they possess, they will have distilled their qubit pairs.
\par
The best-known entanglement distillation protocols use two techniques to gain information about this sequence of Bell states.
Adopting the terminology of Ref.~\cite{HDDM}, we will call these appended ebit measurements (AEMs) and bilateral Pauli measurements (BPMs).
Both of these techniques allow us to perform a parity check.
By this, we mean that if our sequence of Bell states is $\Ket{\Phi_{y_{1}y_{2}}}\dots\Ket{\Phi_{y_{2n-1}y_{2n}}}$, for some $y \in \left\{0, 1\right\}^{2n}$, then, by performing a parity check with respect to some sequence $x \in \left\{0, 1\right\}^{2n}$, we are able to discover the value of
\begin{equation}
    x^{\intercal}y = \sum_{k = 1}^{2n} x_{k}y_{k} \pmod{2}\text{.}
\end{equation}
From now on, all addition and multiplication in such parity checks will be performed element-wise and modulo $2$.
\par
In Ref.~\cite{VDVNDMV}, the authors show that all the permutations of Bell pair product states that Alice and Bob can perform via unitary operations can be performed by Clifford operations.
Up to a complex phase, the Clifford operations on $n$ qubit pairs have a one-to-one correspondence with the symplectic matrices, $S \in \mathbb{Z}_{2}^{2n \times 2n}$, satisfying
\begin{equation}
S^{\intercal}PS = P\text{,}
\end{equation}
where
\begin{equation}
    P =
    \begin{pmatrix}
    0 & 1 & 0 & 0 & 0 & & \dots & & 0 \\
    1 & 0 & 0 & 0 & 0 \\
    0 & 0 & 0 & 1 & 0 & & \dots & & \vdots \\
    0 & 0 & 1 & 0 & 0 \\
    0 & & & \ddots & \ddots & \ddots & & & 0\\
    & & & & 0 & 0 & 1 & 0 & 0\\
    \vdots & & \dots & & 0 & 1 & 0 & 0 & 0 \\
    & & & & 0 & 0 & 0 & 0 & 1 \\
    0 & & \dots & & 0 & 0 & 0 & 1 & 0
    \end{pmatrix}
\end{equation}
is a $2n \times 2n$ matrix.
If $S$ is the $2n \times 2n$ symplectic matrix associated with a particular Clifford operation, $C$, then
\begin{equation}
C\Ket{\Phi_{y_{1}y_{2}}}\dots\Ket{\Phi_{y_{2n-1}y_{2n}}} = \Ket{\Phi_{y'_{1}y'_{2}}}\dots\Ket{\Phi_{y'_{2n-1}y'_{2n}}}\text{,}
\end{equation}
where
\begin{equation}
    y' = Sy\text{.}
\end{equation}
\par
To perform an AEM, Alice and Bob must already possess a perfect Bell pair, $\Ket{\Phi_{00}}$.
They can then perform a Clifford operation on their combined state, $\Ket{\Phi_{y_{1}y_{2}}}\dots\Ket{\Phi_{y_{2n-1}y_{2n}}}\Ket{\Phi_{00}}$, mapping it to $\Ket{\Phi_{y_{1}y_{2}}}\dots\Ket{\Phi_{y_{2n-1}y_{2n}}}\Ket{\Phi_{0x^{\intercal}y}}$.
In other words, the initial state remains unchanged and the appended state stores the parity.
By measuring the qubits of the final pair in their computational bases and classically communicating the measurement outcomes to each other, Alice and Bob can then determine the value of $x^{\intercal}y$.
This comes at the cost of destroying a perfect ebit.
Note, however, that by performing this check over many identical systems, we can reduce the number of perfect ebits which need to be consumed on average to the base-$2$ Shannon entropy of $x^{\intercal}y$.
Throughout this paper, we will assume that AEMs are performed in this asymptotic manner.
For a more detailed discussion of this technique, see Ref.~\cite{BBPSSW}.
\par
Alternatively, we can choose a symplectic $2n \times 2n$ matrix whose final row is $x^{\intercal}$.
If Alice and Bob then apply the associated Clifford operation before measuring the qubits of their last pair in their computational bases, they can determine $x^{\intercal}y$.
This is a BPM and comes at the cost of one of the $n$ pairs they are trying to distil.
Note that, by measuring their final pair, Alice and Bob destroy all information about the parity of $y$ with respect to any sequence, $x' \in \left\{0, 1\right\}^{2n}$, satisfying $x'^{\intercal}Px = 1$.
Therefore, subsequent parity checks cannot be with respect to such $x'$ but all other parity checks are still possible.
\par
Given that a BPM consumes a whole pair, unlike an AEM which consumes at most one ebit and usually consumes less, and a BPM restricts the checks we can perform in the future, it might seem that BPMs provide no advantage over AEMs.
However, note that the measurement result and resulting state would have been the same if our sequence of states had been characterized by $y + Px$ instead of $y$.
Therefore, after performing the BPM, there is no need for us to distinguish between the state $y$ and the state $y + Px$.
The potential advantage of a BPM over an AEM results entirely from this.

\section{Updating the Parity Check Shortlists}\label{UpdatingShortlists}

Suppose that we want to perform an AEM or a BPM on our system and we have to choose between performing an AEM with respect to some parity check in $A$ or a BPM with respect to some parity check in $B$, where $A$ and $B$ are subspaces of $\left\{0, 1\right\}^{2n}$, which is thought of as a vector space under element-wise modulo $2$ addition.
Note that, although the zero parity check will be in both $A$ and $B$, we will assume that this is never chosen because there is never any information to be gained by performing this parity check.
Let us choose $(\tilde{a}_{1}, \tilde{a}_{2}, \dots, \tilde{a}_{m_{A}})$ and $(\tilde{b}_{1}, \tilde{b}_{2}, \dots, \tilde{b}_{m_{B}})$ to be bases for $A$ and $B$, where $m_{A}$ and $m_{B}$ are the dimensionalities of $A$ and $B$, respectively.
After we have performed a check, whether by AEM or BPM, there will be certain elements of $A$ and $B$ that we can remove from consideration for future checks.
\par
In this section, we will describe a method of updating our bases for these shortlists.
In this way, we will be able to update both shortlists so that there is no loss of generality in the AEMs and BPMs we consider and it will be possible to perform any parity check remaining in either shortlist.
The process is also described in Algorithm~\ref{alg:basisUpdate}.

\begin{widetext}

\noindent\rule{\textwidth}{0.8pt}

\algorithmcaption{Dynamic Basis Updating for Measurement Shortlists}
\label{alg:basisUpdate}

\noindent\rule{\textwidth}{0.4pt}

\begin{algorithmic}[1]

\Require Performed check $c$, check type
$T \in \{\text{AEM}, \text{BPM}\}$,
AEM basis $\tilde{A}$, BPM basis $\tilde{B}$,
parity operator $P$.

\Ensure Updated bases $\tilde{A}'$ and $\tilde{B}'$.

\vspace{0.2cm}

\Procedure{RemoveRedundancy}{$c, L$}
    \Comment{Implementation of function $f$}

    \State Let
    $L = (\tilde{c}_1, \tilde{c}_2, \dots, \tilde{c}_m)$

    \If{$\exists x \in \{0,1\}^{m}\setminus\{\mathbf{0}\}$
    such that
    $c = \sum_{k=1}^{m} x_k \tilde{c}_k$}

        \State
        $k_f \gets
        \min\{k \in \{1,2,\dots,m\}\mid x_k=1\}$

        \State \Return
        $(\tilde{c}_1,\dots,
        \tilde{c}_{k_f-1},
        \tilde{c}_{k_f+1},
        \dots,\tilde{c}_m)$

    \Else

        \State \Return $L$

    \EndIf

\EndProcedure

\vspace{0.2cm}

\Procedure{EnforceCommutativity}{$c, L$}
    \Comment{Implementation of function $g$}

    \State Let
    $L = (\tilde{c}_1, \tilde{c}_2, \dots, \tilde{c}_m)$

    \If{$\exists k \in \{1,2,\dots,m\}$
    such that
    $\tilde{c}_k^{\intercal}Pc=1$}

        \State
        $k_g \gets
        \min\{
            k \in \{1,2,\dots,m\}
            \mid
            \tilde{c}_k^{\intercal}Pc=1
        \}$

        \State
        $L' \gets \emptyset$
        \Comment{Initialize empty updated list}

        \For{$k \in
        \{1,2,\dots,m\}\setminus\{k_g\}$}

            \State
            $\tilde{c}'_k
            \gets
            \tilde{c}_k
            +
            (\tilde{c}_k^{\intercal}Pc)
            \tilde{c}_{k_g}$

            \State Append $\tilde{c}'_k$ to $L'$

        \EndFor

        \State \Return $L'$

    \Else

        \State \Return $L$

    \EndIf

\EndProcedure

\vspace{0.2cm}

\Procedure{UpdateShortlists}
{$c, T, \tilde{A}, \tilde{B}$}

    \If{$T=\text{AEM}$}

        \State
        $\tilde{A}'
        \gets
        \Call{RemoveRedundancy}
        {c,\tilde{A}}$

        \State
        $\tilde{B}'
        \gets
        \Call{EnforceCommutativity}
        {c,\tilde{B}}$

    \ElsIf{$T=\text{BPM}$}

        \State
        $\tilde{A}_{\text{temp}}
        \gets
        \Call{RemoveRedundancy}
        {c,\tilde{A}}$

        \State
        $\tilde{A}'
        \gets
        \Call{EnforceCommutativity}
        {c,\tilde{A}_{\text{temp}}}$

        \State
        $\tilde{B}_{\text{temp}}
        \gets
        \Call{RemoveRedundancy}
        {c,\tilde{B}}$

        \State
        $\tilde{B}'
        \gets
        \Call{EnforceCommutativity}
        {c,\tilde{B}_{\text{temp}}}$

    \EndIf

    \State \Return $\tilde{A}',\tilde{B}'$

\EndProcedure

\end{algorithmic}

\vspace{-0.2cm}
\noindent\rule{\textwidth}{0.8pt}

\end{widetext}




\par
To help us with our description, we first define two functions, $f$ and $g$.
For every $n' \in \mathbb{N}$, let $O_{n'}$ denote the set of ordered lists of linearly independent elements of $\left\{0, 1\right\}^{2n'}$.
Then both $f$ and $g$ are functions from $\cup_{n' \in \mathbb{N}}{\left(\left\{0, 1\right\}^{2n'} \times O_{n'}\right)}$ to $\cup_{n'\in\mathbb{N}}{O_{n'}}$.
\par
For some $\left(m, n'\right) \in \mathbb{N}^{2}$, let $c$  be in $\left\{0, 1\right\}^{2n'}$ and $\left(\tilde{c}_{1}, \tilde{c}_{2}, \dots, \tilde{c}_{m}\right)$ be in $O_{n'}$.
Then, if there is a non-zero $x$ in $\left\{0, 1\right\}^{m}$ such that $c = \sum_{k=1}^{m}x_{k}\tilde{c}_{k}$, this $x$ will be unique and we can let
\begin{equation}
k_{f} = \min{\left\{k \in \left\{1, 2, \dots, m\right\} \mid x_{k} = 1\right\}}
\end{equation}
and $f\left(\left(c, \tilde{c}_{1}, \tilde{c}_{2}, \dots, \tilde{c}_{m}\right)\right)$ be 
\begin{equation}
     \left(\tilde{c}_{1}, \dots, \tilde{c}_{k_{f}-1}, \tilde{c}_{k_{f}+1}, \dots, \tilde{c}_{m}\right)\text{.}
\end{equation}
If, however, no such $x$ exists, then we just let
\begin{equation}
f\left(\left(c, \tilde{c}_{1}, \tilde{c}_{2}, \dots, \tilde{c}_{m}\right)\right) = \left(\tilde{c}_{1}, \tilde{c}_{2}, \dots, \tilde{c}_{m}\right)\text{.}
\end{equation}
Likewise, if, for some $k \in \left\{1, 2, \dots, m\right\}$, we have
\begin{equation}
    \tilde{c}_{k}^{\intercal}Pc = 1\text{,}
\end{equation}
then we can let
\begin{equation}
    k_{g} = \min\left\{k \in \left\{1, 2, \dots, m\right\} \mid \tilde{c}_{k}^{\intercal}Pc = 1\right\}\text{,}
\end{equation}
\begin{equation}
\tilde{c}'_{k} = \tilde{c}_{k} + \tilde{c}_{k}^{\intercal}Pc\tilde{c}_{k_{g}}
\end{equation}
for $k \in \left\{1, 2, \dots, m\right\} \setminus \left\{k_{g}\right\}$, and set $g\left(\left(c, \tilde{c}_{1}, \tilde{c}_{2}, \dots, \tilde{c}_{m}\right)\right)$ equal to
\begin{equation}
      \left(\tilde{c}'_{1}, \tilde{c}'_{2}, \dots, \tilde{c}'_{k_{g}-1}, \tilde{c}'_{k_{g}+1}, \dots, \tilde{c}'_{m}\right)\text{.}
\end{equation}
If no such $k$ exists, we simply set
\begin{equation}
    g\left(\left(c, \tilde{c}_{1}, \tilde{c}_{2}, \dots, \tilde{c}_{m}\right)\right) = \left(\tilde{c}_{1}, \tilde{c}_{2}, \dots, \tilde{c}_{m}\right)\text{.}
\end{equation}
\par
Suppose that we perform an AEM with respect to some $a \in A$.
Once we learn the result, for any $a' \in A$, performing an AEM with respect to $a'$ will be the same as performing an AEM with respect to $a' + a$.
Therefore, we can halve the size of our shortlist of AEMs without loss of generality by keeping only one of every pair of parity checks in $A$ separated by the addition of $a$.
In particular, we replace our basis, $\left(\tilde{a}_{1}, \tilde{a}_{2}, \dots, \tilde{a}_{m_{A}}\right)$, with
\begin{equation}
    f\left(\left(a, \tilde{a}_{1}, \tilde{a}_{2}, \dots, \tilde{a}_{m_{A}}\right)\right)\text{.}
\end{equation}
\par
Now, note that, after performing this AEM, we will only be interested in performing a BPM with respect to some $b \in B$ if $a^{\intercal}Pb = 0$.
Otherwise, the BPM loses its only possible advantage over the equivalent AEM, namely that it maps states represented by vectors differing by the addition of $Pb$ to the same state, because only one state in each pair would be consistent with the AEM with respect to $a$.
Therefore, we replace our basis for $B$, $\left(\tilde{b}_{1}, \tilde{b}_{2}, \dots, \tilde{b}_{m_{B}}\right)$, with
\begin{equation}
    g\left(\left(a, \tilde{b}_{1}, \tilde{b}_{2}, \dots, \tilde{b}_{m_{B}}\right)\right)\text{.}
\end{equation}
Although after applying our AEM the result of a BPM with respect to some $b \in B$ would be determined by the result of a BPM with respect to $b + a$, we do not update our shortlist of BPMs in the way we did our shortlist of AEMs using $f$ because the states that are made indistinguishable by a BPM with respect to $b$ are not the same as the states that are made indistinguishable by a BPM with respect to $b + a$.
\par
Now let us suppose that, instead of performing an AEM with respect to $a$, we had performed a BPM with respect to some $b \in B$.
We must make sure that all future parity checks, whether they be AEMs or BPMs, are consistent with this check.
That is, if we want to perform a parity check with respect to $x'$, we must ensure that $x'^{\intercal}Pb = 0$.
\par
To update our shortlist of AEMs after our BPM, we can use both of our techniques.
Firstly, we note that, having performed our BPM with respect to $b$, performing an AEM with respect to some $a$ in $A$ will be equivalent to performing an AEM with respect to $a + b$.
By applying $f$ appropriately, we can remove these redundant checks.
Secondly, as discussed above, there are some checks that we must remove and these can be removed using $g$.
In particular, our updated basis will be
\begin{equation}
    g\left(\left(b\right)\frown f\left(\left(b, \tilde{a}_{1}, \tilde{a}_{2}, \dots, \tilde{a}_{m_{A}}\right)\right)\right)\text{,}
\end{equation}
where $\frown$ denotes the concatenation of lists.
\par
Likewise, we can employ both techniques to update our shortlist of BPMs after performing our BPM with respect to $b$.
Unlike in the aftermath of an AEM, there is no loss of generality in using $f$ to update our BPM shortlist.
Indeed, after performing a BPM with respect to $b$, the states that are made indistinguishable by a further BPM with respect to an allowed parity check, $b'$, in $B$ are the same as those that are made indistinguishable by a BPM with respect to $b' + b$ because states differing by $Pb$ will already have been made indistinguishable.
Therefore, our updated basis for the BPM shortlist will be
\begin{equation}
    g\left(\left(b\right)\frown f\left(\left(b, \tilde{b}_{1}, \tilde{b}_{2}, \dots, \tilde{b}_{m_{B}}\right)\right)\right)\text{.}
\end{equation}

\section{The Receding Horizon Search}\label{RecedingHorizonSearch}
We have, then, established a framework for distilling entangled qubit pairs via AEMs and BPMs.
We choose a parity check from one of our two shortlists, apply it, update our shortlists as described in Sec.~\ref{UpdatingShortlists} and repeat the process until we have distilled all remaining pairs.
The specification of a protocol now lies in how we choose which AEM or BPM to perform at each stage.
\par
Suppose we are part way through this process.
Through our AEMs and BPMs, we will have gained information about which sequence of Bell states our system was originally composed of.
Moreover, our BPMs will have rendered the distinction between some of these sequences irrelevant, partitioning the set of possible sequences into equivalence classes of all sequences which can be considered identical.
We will let $\sigma$ denote the resulting probability distribution over this set of equivalence classes and we will let $m$ denote the number of qubit pairs we still have.
The number of our equivalence classes will be $4^{m}$.
Finally, we will let $\tilde{A}$ and $\tilde{B}$ denote respectively our bases for $A$ and $B$, our shortlists of AEMs and BPMs.

\par
Now, if $\sigma$'s support contains multiple equivalence classes, we are not yet finished and, to complete the distillation, we must perform more AEMs and BPMs.
If we had specified how we would choose which AEMs and BPMs to perform, we could call the average number of additional BPMs we would perform plus the average number of additional perfect ebits we would consume to complete the distillation the parity check count, denoted $C\left(\sigma, \tilde{A}, \tilde{B}\right)$.
The expected number of surplus ebits we would distil would then be
\begin{equation}
    m - C\left(\sigma, \tilde{A}, \tilde{B}\right)\text{.}
\end{equation}
Our expected number of surplus ebits from the beginning would be
\begin{equation}
    n - C\left(\left(\rho_{00}, \rho_{01}, \rho_{10}, \rho_{11}\right)^{\otimes n}, \tilde{K}_{n}, \tilde{K}_{n}\right)\text{,}
\end{equation}
where, for every $l \in \mathbb{N}$,
\begin{equation}
\tilde{K}_{l} = \left(\left(1, 0, \dots, 0\right), \left(0, 1, \dots, 0\right), \dots, \left(0, 0, \dots, 1\right)\right)\text{,}
\end{equation}
a sequence of $2l$ elements of $\left\{0, 1\right\}^{2l}$, is a basis for both of our initial shortlists.
The average yield would be
\begin{equation}
    \frac{n - C\left(\left(\rho_{00}, \rho_{01}, \rho_{10}, \rho_{11}\right)^{\otimes n}, \tilde{K}_{n}, \tilde{K}_{n}\right)}{n}\text{.}
\end{equation}
Our aim is to find a way of choosing AEMs and BPMs that keeps the the parity check count as low as we can.
\par
Our protocol will have two parameters, $\left(d, r\right) \in \mathbb{N}^{2}.$
If we ever have $m = 1$ and $r > 1$, we join $r$ identical and independent copies of our system together before performing any more parity checks.
To choose which AEM or BPM to perform, we will search through the tree of possible sequences of parity checks from our shortlists and measurement results to a search depth $d$.
That is, we will consider all such sequences up to the point where our system has been distilled or we have performed $d$ measurements.
Then, for every leaf node in our tree, we will combine the average contribution to the parity check count of the AEMs and BPMs along the path from the root node to that leaf node with an estimate of the additional parity check count we will incur in completing the distillation from the leaf node.
In this way, we can associate an estimated parity check count with every possible sequence of parity checks and measurement results at depth $d$.
Then, weighting these estimates by the probabilities of different measurement outcomes, we can choose an AEM or BPM that minimizes our average estimated parity check count, assuming we will always choose the AEM or BPM that minimizes our estimated average parity check count until we reach a leaf node. We will now make this process more precise.



\par
If $x$ is in $A \cup B$, then we will let $\sigma_{x}$ be parity of our state with respect to $x$, treated as a random variable.
Furthermore, if $a$ is in $A$, we will let $X_{\sigma, a, 0}$ denote the distribution over equivalence classes that results if we perform an AEM with respect to $a$ on $\sigma$ and find that the parity is $0$ and $X_{\sigma, a, 1}$ the distribution that results if we perform the same parity check and find that the parity is $1$.
Likewise, if $b$ is in $B$, we will let $Y_{\sigma, b, 0}$ and $Y_{\sigma, b, 1}$ denote the corresponding probability distributions for a BPM with respect to $b$.
Strictly speaking, some parity results occur with probability $0$ when we perform certain parity checks on certain states.
For the sake of making $X$ and $Y$ well-defined, we will take the resulting probability distributions to be uniform, with the number of remaining qubit pairs being unaffected by $X$ and reduced by $1$ by $Y$.
\par
Given $a$ in $A$ or $b$ in $B$, we would like to define a quantity, $D\left(\sigma, \tilde{A}, \tilde{B}, a, d\right)$ or $E\left(\sigma, \tilde{A}, \tilde{B}, b, d\right)$, that acts as an estimate for the parity check count we can expect to achieve from our state $\sigma$ if we now apply an AEM with respect to $a$ or a BPM with respect to $b$, respectively.
We first note that we can always achieve a parity check count of $H\left(\sigma\right)$, where $H$ denote the base-2 Shannon entropy, by performing AEMs.
Likewise, we can always achieve a parity check count of $m$ by applying BPMs.
Based on this, for every $a \in A$ and $b \in B$, we let $D\left(\sigma, \tilde{A}, \tilde{B}, a, 0\right)$ and $E\left(\sigma, \tilde{A}, \tilde{B}, b, 0\right)$ both equal
\begin{equation}
    H^{\star}\left(\sigma\right) = \min\left(\left\{m, H\left(\sigma\right)\right\}\right)\text{.}\label{BaseEvaluation}
\end{equation}
Then, if $m > 1$ or $r = 1$, for every $d \in \mathbb{N}$, $a \in A$ and $b \in B$, we let $D\left(\sigma, \tilde{A}, \tilde{B}, a, d\right)$ be 
\begin{equation}
     H\left(\sigma_{a}\right) + \sum_{i=0}^{1}{\Pr\left(\sigma_{a}=i\right)\tilde{C}\left(X_{\sigma, a, i}, T, U, d-1\right)}\label{AEMEvaluation}
\end{equation}
and $E\left(\sigma, \tilde{A}, \tilde{B}, b, d\right)$ be
\begin{equation}
     1 + \sum_{i=0}^{1}{\Pr\left(\sigma_{b}=i\right)\tilde{C}\left(Y_{\sigma, b, i}, V, W, d-1\right)}\text{,}\label{BPMEvaluation}
\end{equation}
where
\begin{equation}
    T = f\left(\left(a\right)\frown\tilde{A}\right)\text{,}
\end{equation}
\begin{equation}
    U = g\left(\left(a\right)\frown \tilde{B}\right)\text{,}
\end{equation}
\begin{equation}
    V = g\left(\left(b\right)\frown f\left(\left(b\right) \frown \tilde{A}\right)\right)\text{,}
\end{equation}
\begin{equation}
    W = g\left(\left(b\right)\frown f\left(\left(b\right) \frown \tilde{B}\right)\right)
\end{equation}
and, for every distribution over equivalence classes, $Z$, AEM shortlist basis, $\tilde{A}'$, and BPM shortlist basis, $\tilde{B}'$, that might result from our parity check, $\tilde{C}\left(Z, \tilde{A}', \tilde{B}', d-1\right)$ is
\begin{equation}
    \min{\left(L_{A}\cup L_{B}\right)}\text{,}
\end{equation}
where
\begin{equation}
    L_{A} = \cup_{a \in \text{span}\left(\tilde{A}'\right)}{\left\{D\left(Z, \tilde{A}', \tilde{B}', a, d-1\right)\right\}}
\end{equation}
and
\begin{equation}
    L_{B} = \cup_{b \in \text{span}\left(\tilde{B}'\right)}{\left\{E\left(Z, \tilde{A}', \tilde{B}', b, d-1\right)\right\}}\text{.}
\end{equation}
If $m = 1$ and $r > 1$, then, for every such $Z$, $\tilde{A}'$ and $\tilde{B}'$ and $d \in \left\{0, 1, 2, \dots\right\}$, we let
\begin{equation}
    \tilde{C}\left(Z, \tilde{A}', \tilde{B}', d\right) = \frac{\tilde{C}\left(Z^{\otimes r}, \tilde{K}_{r}, \tilde{K}_{r}, d\right)}{r}\text{.}
\end{equation}
\par
\begin{figure}
    \centering
    \includegraphics[width=0.48\textwidth]{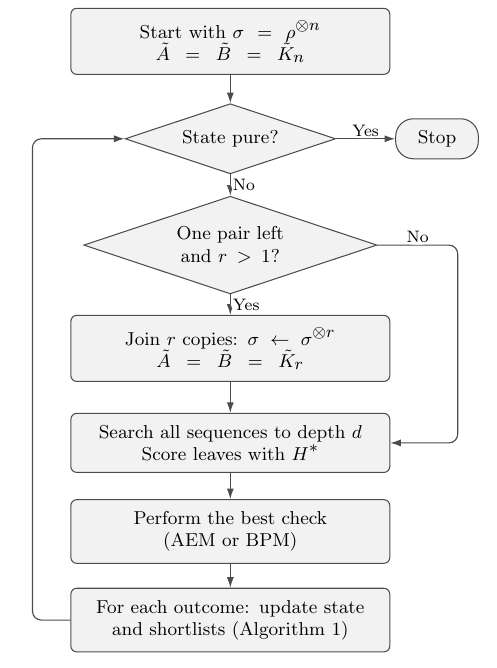}
    \caption{Our approach to selecting parity checks.}
    \label{fig:Flowchart}
\end{figure}

Using our functions $D$ and $E$, we can estimate the parity check count that we will achieve if we start by performing an AEM with respect to some $a \in A$ or by performing a BPM with respect to some $b \in B$.
At each stage, then, we apply the parity check that we estimate will allow us to achieve the lowest parity check count and therefore the highest yield.
This process is summarized in Fig.\ref{fig:Flowchart}.
For the exact details of the implementation used to produce our figures, including how we dealt with different parity checks being assigned equal estimated value by $D$ and $E$, please see Appendix~\ref{Psuedocode}.
\par
Finally, note that for every distribution over equivalence classes, $Z$, AEM shortlist basis, $\tilde{A}'$, BPM shortlist basis, $\tilde{B}'$, and $d' \in \left\{0, 1, 2, \dots, d\right\}$ we will have
\begin{equation}
    \tilde{C}\left(Z, \tilde{A}', \tilde{B}', d'\right) \geq C\left(Z, \tilde{A}', \tilde{B}'\right)\text{.}
\end{equation}
That is, our estimates actually provide upper bounds on the true parity check count.
This is because our protocol always pursues the lowest parity check count that our search has shown to be achievable and, as we continue with the protocol, this lowest value can only decrease.

\section{Applying the Receding Horizon Search}\label{Applications}

\subsection{Distilling Qubit Werner States}

Let us calculate the parity check count of the protocol of Sec.~\ref{RecedingHorizonSearch} when applied to a particular example.
Naturally, if our state has already been distilled, then we have
\begin{equation}
    C\left(\sigma, \tilde{A}, \tilde{B}\right) = 0\text{.}
\end{equation}
If $m = 1$ and $r > 1$, then
\begin{equation}
    C\left(Z, \tilde{A}', \tilde{B}'\right) = \frac{C\left(Z^{\otimes r}, \tilde{K}_{r}, \tilde{K}_{r}\right)}{r}\text{.}
\end{equation}
Otherwise, if the next parity check selected based on the technique of Sec.~\ref{RecedingHorizonSearch} is an AEM with respect to some $a \in A$, then $C\left(\sigma, \tilde{A}, \tilde{B}\right)$ will be
\begin{equation}
      H\left(\sigma_{a}\right) + \sum_{i=0}^{1}{\Pr\left(\sigma_{a}=i\right)C\left(X_{\sigma, a, i}, T, U\right)}\label{AEMRecursion}
\end{equation}
and, if we select a BPM with respect to some $b \in B$, then it will be
\begin{equation}
     1 + \sum_{i=0}^{1}{\Pr\left(\sigma_{b}=i\right)C\left(Y_{\sigma, b, i}, V, W\right)}\text{.}\label{BPMRecursion}
\end{equation}
In this way, we can calculate the yield of our protocol.
\par
\begin{figure}[ht]
\centering
\vspace{-0.2cm}
\includegraphics[width=0.48\textwidth]{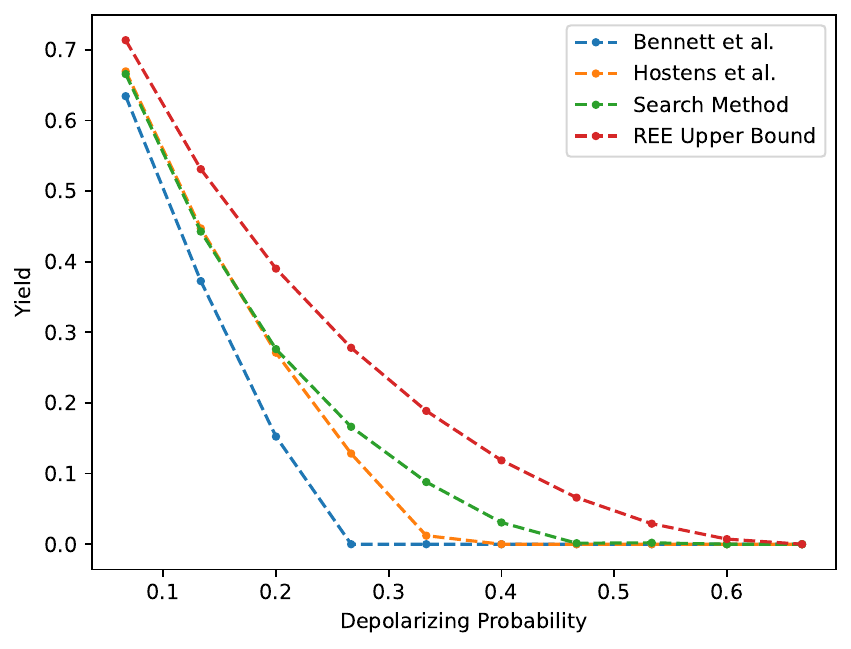}
\caption{The average yield of our search-based protocol (green), compared to that of the asymptotic protocols of Ref.~\cite{BBPSSW} (blue) and Ref.~\cite{HDDM} (orange), when applied to the Choi state of a qubit depolarising channel. The receding horizon search was conducted with $\left(n, r, d\right) = \left(4, 4, 3\right)$. In the language of Ref.~\cite{HDDM}, we used $q = 6$ for their protocol or, in our language, $n = 64$. An upper bound on the achievable yield, the relative entropy of entanglement \cite{VPRK} of $\rho$, is shown in red. The dotted lines are linear interpolations between neighbouring points. The plots are based on exact numerical evaluations, not stochastic simulations.}
\label{fig:NoDEJMPSDepolarising}
\end{figure}

In Fig.~\ref{fig:NoDEJMPSDepolarising}, we compare the performance of our protocol with that of the asymptotic protocols of Refs.~\cite{BBPSSW} and~\cite{HDDM} when applied to Werner states with Bell fidelity $F \in \left\{0.5, 0.55, 0.6, 0.65, \dots, 0.95\right\}$, states of the form
\begin{align}
    \rho &= W_{F} \nonumber \\
    &= \frac{\left(4F-1\right)\Ket{\Phi_{00}}\Bra{\Phi_{00}} + \left(1-F\right)\mathds{1}_{4}}{3}\text{,}
\end{align}
where $\mathds{1}_{4}$ is the identity operator on our qubit pair's Hilbert space.
Such states are the Choi states of qubit depolarising channels, with the depolarising probability, $p$, being given by
\begin{equation}
    p = \frac{4\left(1 - F\right)}{3}\label{DepolarisingProbability}
\end{equation}
or
\begin{equation}
    F = \frac{4 - 3p}{4}\text{.}
\end{equation}
Therefore, the yields achieved by our protocol provide lower bounds on the two-way-assisted quantum capacity of qubit depolarising channels~\cite{pirandola2017fundamental}.
In Fig.~\ref{fig:NoDEJMPSDepolarising}, we can see that the performance our protocol is better than that of Ref.~\cite{HDDM} for a wide range of values of $F$.

\begin{figure}[ht]
\centering
\vspace{-0.2cm}
\includegraphics[width=0.48\textwidth]{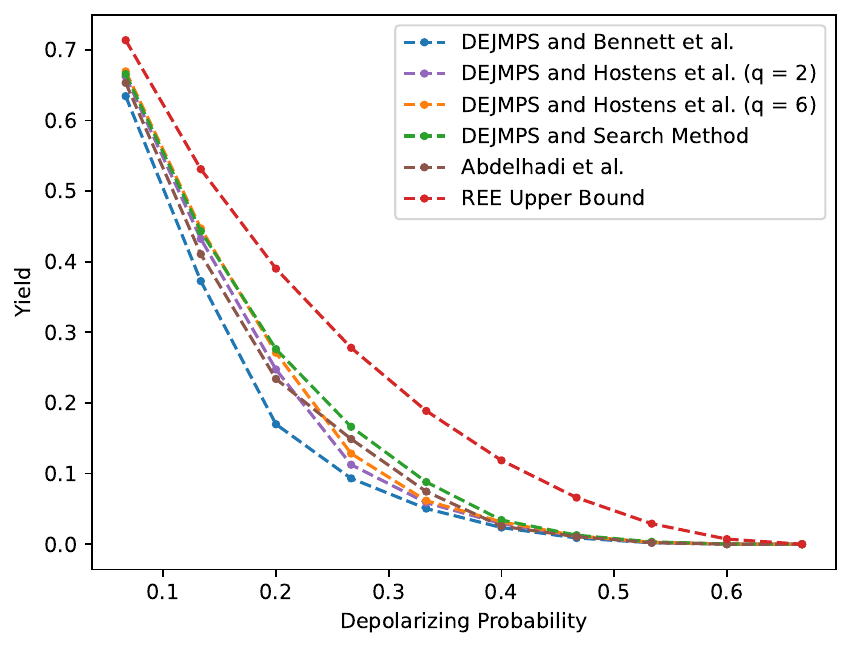}
\caption{The average yield of our search-based protocol (green), compared to that of the asymptotic protocols of Ref.~\cite{BBPSSW} (blue), Ref.~\cite{AJOCSS} (brown) and Ref.~\cite{HDDM} (purple and orange), when applied to the output of a qubit depolarising channel. Each application of an asymptotic protocol was preceded by an optimal number of iterations of the DEJMPS protocol. The results of the receding horizon protocol were found using $\left(n, r, d\right) = \left(4, 4, 3\right)$. In the language of Ref.~\cite{HDDM}, we used $q = 2$ and $q = 6$ for their protocol or, in our language, $n = 4$ and $n = 64$. An upper bound on the achievable yield, the relative entropy of entanglement Ref.~\cite{VPRK} of $\rho$, is shown in red. The dotted lines are linear interpolations between neighbouring points. The plots are based on exact numerical evaluations, not stochastic simulations.}
\label{fig:Depolarising}
\end{figure}

It is well-known that the yields of the protocols of Refs.~\cite{BBPSSW} and~\cite{HDDM} can be improved by applications of the recurrence protocol of Ref.~\cite{DEJMPS}, at least when $F$ is sufficiently small.
Likewise, the protocol of Ref.~\cite{AJOCSS} begins with applications of this recurrence protocol.
We found that the yield of our protocol could be similarly improved.
Fig.~\ref{fig:Depolarising} shows the performance of the asymptotic protocols of Refs.~\cite{AJOCSS,BBPSSW,HDDM} and our own protocol, applied to the same states as in Fig.~\ref{fig:NoDEJMPSDepolarising}, when preceded by an optimal number of iterations of this recurrence protocol.
Before applying an iteration of the recurrence protocol, we always permuted the Bell states of each pair to maximize the increase in the Bell fidelity upon success.
Before applying the protocol of Ref.~\cite{HDDM}, we always permuted the Bell states of each pair to maximize the yield.
A similar permutation is part of the protocol of Ref.~\cite{AJOCSS}.
We see that our protocol outperforms that of Ref.~\cite{AJOCSS} and that of Ref.~\cite{HDDM} with $q = 2$ for all values of $F$ tested and that of Ref.~\cite{HDDM} with $q = 6$ for all values of $F$ tested except $0.9$ and $0.95$.

\par
In principle, the search technique described in Sec.~\ref{RecedingHorizonSearch} might have discovered the recurrence protocol of Ref.~\cite{DEJMPS} on its own.
Indeed, for $F \in \left\{0.75, 0.8, 0.85, 0.9, 0.95\right\}$, we found that the optimal number of iterations of the recurrence protocol to perform before applying our protocol was $0$, unlike the protocol of Ref.~\cite{HDDM}, for which this only occurred for $F \in \left\{0.8, 0.85, 0.9, 0.95\right\}$, and the protocol of Ref.~\cite{BBPSSW}, for which this only occurred for $F \in \left\{0.9, 0.95\right\}$.
However, for lower values of $F$, assisting the search by applying these recurrence iterations improved our yield as $d$ was not large enough for the utility of the recurrence protocols to become clear in the search algorithm.
\par
Of course, the performance of our protocol depends on our choice of $n$, $r$ and $d$.
Our results shown in Fig.~\ref{fig:Depolarising} were found using $\left(n, r, d\right) = \left(4, 4, 3\right)$.
By comparison, in our language, the protocols of Refs.~\cite{AJOCSS,HDDM} had $\left(n, r\right)$ equal to $\left(4, 2\right)$ and $\left(64, 1\right)$ respectively.
Indeed, the superior performance of the protocol of Ref.~\cite{HDDM} for high $F$ is not observed when we set $q = 2$, suggesting it can be explained by the larger value of $n$ used.
Certainly, we would achieve at least as strong performance as that of the protocol of Ref.~\cite{HDDM} using the technique of Sec.~\ref{RecedingHorizonSearch} with $\left(n, r, d\right) = \left(64, 1, 128\right)$, although actually performing such a search would be wholly impractical.
The selection of our protocol parameters is discussed in more detail in Section~\ref{Parameters}.

\subsection{Protocol Parameter Selection}\label{Parameters}

The protocol described in Sec.~\ref{RecedingHorizonSearch} depends on three parameters, $n$, $r$ and $d$.
Our choice of these parameters has a significant effect on the resulting yield of the protocol and the time it takes to carry out Algorithm~\ref{alg:distillation_search}.
Here, we will provide some discussion on the selection of these parameters.
\par
Increasing $n$ allows us to search over longer parity check vectors, which measure correlations between more pairs.
Being able to measure correlations between a large number of pairs could be important for certain states.
For instance, if our state contains little noise, there is little information to be gained on average by any parity check on a small number of copies of the state.
This would make it difficult to find BPMs worth performing since a BPM sacrifices a whole pair and, to beat the asymptotic protocol of Ref.\cite{BBPSSW}, we must find such BPMs.
\par
However, there are disadvantages to increasing $n$ apart from the increased computational resources that will then be required to perform the search.
One of these relates to how $n$ can or cannot be factorized into smaller numbers of qubit pairs.
For example, suppose that an effective protocol when applied to some state is to perform one iteration of the recurrence protocol of Ref.~\cite{DEJMPS} and, if that iteration succeeds, to apply the asymptotic protocol of Ref.~\cite{BBPSSW}.
If we had $n=2$, then our search could consider this possibility.
If we instead had $n=3$, then this protocol would not be considered because the protocol of Ref.~\cite{DEJMPS} is always applied to $2$ qubit pairs at a time and $3$ is not even.
However, this effect should be small, particularly for large $n$. 
A more serious disadvantage is that increasing $n$ tends to effectively decrease our search depth.
To see this, note that for any $\left(n_{0}, k, d_{0}\right) \in \mathbb{N}^{3}$, if we were to search through the tree of parity checks with $n = kn_{0}$ and $r=1$, then to find a sequence of parity checks equivalent to one we could find with $\left(n, r, d\right) = \left(n_{0}, 1, d_{0}\right)$ , we would have to set $d = kd_{0}$.
This is because, instead of considering sequences of $d_{0}$ parity checks on $n_{0}$ qubit pairs, we would have to consider sequences of $kd_{0}$ parity checks, $d_{0}$ on each of the $k$ collections of $n_{0}$ qubit pairs. 
In other words, we could argue that the effective depth of a search is
\begin{equation}
    d_{\text{effective}} = \frac{d}{n}\text{.}
\end{equation}
\par
\begin{figure}[h!]
\centering
\vspace{0.1cm}
\includegraphics[width=0.48\textwidth]{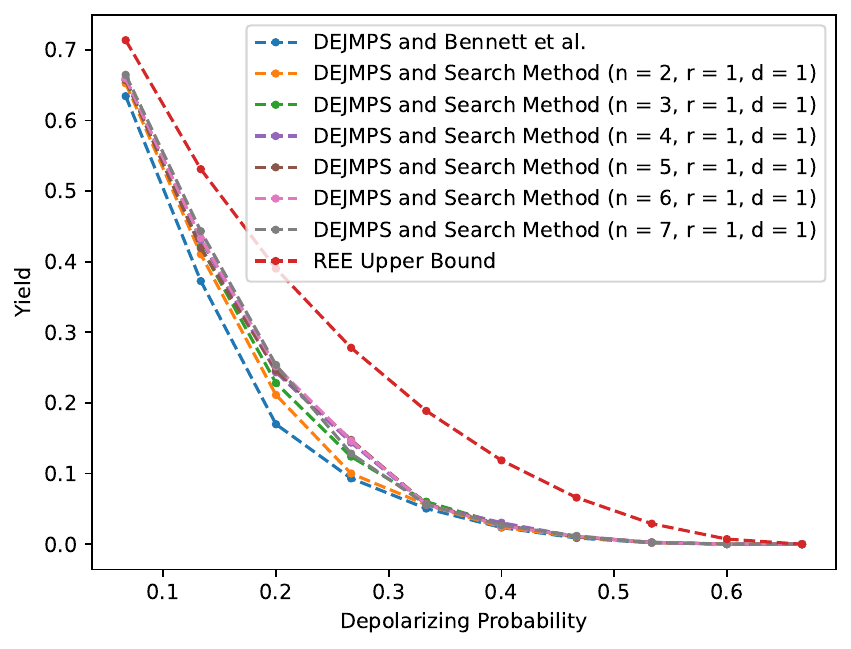}
\caption{The average yield of our search-based protocol with $\left(r, d\right) = \left(1, 1\right)$ and $n \in \left\{2, 3, \dots, 6, 7\right\}$ when applied to Werner states with Bell fidelity in $\left\{0.5, 0.55, 0.6, 0.65, \dots, 0.95\right\}$. The performance of the asymptotic protocol of Ref.~\cite{BBPSSW}, which is equivalent to our protocol with $\left(n, r\right) = \left(1, 1\right)$, is also shown. All asymptotic protocols are preceded by an optimal number of iterations of the protocol of Ref.~\cite{DEJMPS}. An upper bound, the relative entropy of entanglement of the state to be distilled is shown in red. The dotted lines interpolate linearly between neighbouring points. The value of the depolarising probability on the $x$-axis is given by Eq.~\eqref{DepolarisingProbability}.}
\label{fig:VaryingBreadth}
\end{figure}

In Fig.~\ref{fig:VaryingBreadth}, we see the performance of our protocol for different values of $n$ with $\left(r, d\right) = \left(1, 1\right)$.
Note that the asymptotic protocol of Ref.~\cite{BBPSSW} is equivalent to the protocol of Sec.~\ref{RecedingHorizonSearch} with $\left(n, r\right) = \left(1, 1\right)$ because with $\left(n, r\right) = \left(1, 1\right)$ we can only perform AEMs or perform a single BPM to destroy our only qubit pair.
In the figure, there is no monotonic relationship between $n$ and the yield of the protocol, although larger values of $n$ generally lead to higher yields, especially for lower Bell fidelities, with the worst performance always seen by the protocol of Ref.~\cite{BBPSSW} and the best performance coming from the protocol with $n = 7$ when the Bell fidelity is $0.9$ or $0.95$.
This is in accordance with the discussion above.

\par
\begin{figure}[h!]
\vspace{-0.2cm}
\centering
\includegraphics[width=0.48\textwidth]{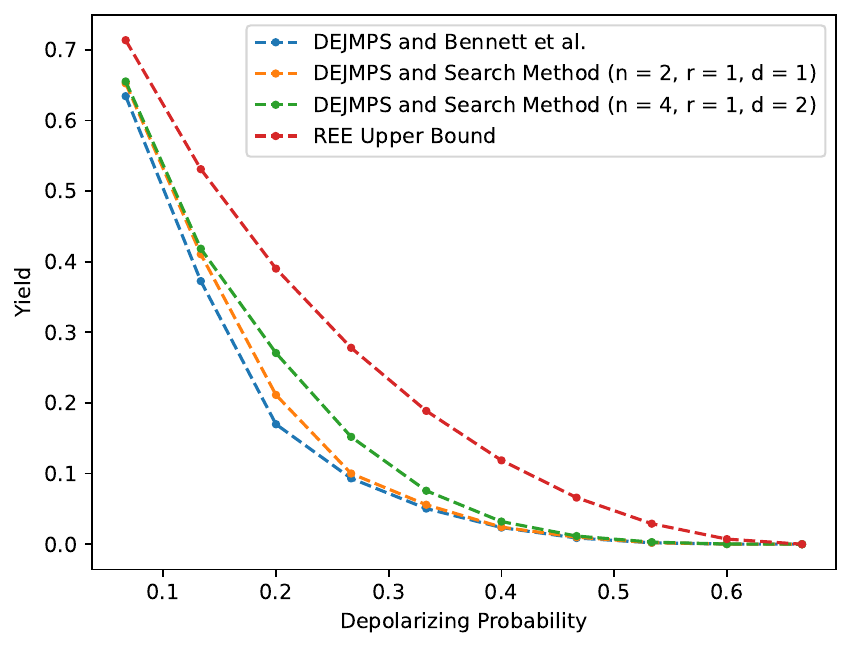}
\caption{The average yield of our search-based protocol with $\left(n, r, d\right) \in \left\{\left(2, 1, 1\right), \left(4, 1, 2\right)\right\}$ when applied to Werner states with Bell fidelity in $\left\{0.5, 0.55, 0.6, 0.65, \dots, 0.95\right\}$. The performance of the asymptotic protocol of Ref.~\cite{BBPSSW} is also shown. All asymptotic protocols are preceded by an optimal number of iterations of the protocol of Ref.~\cite{DEJMPS}. An upper bound, the relative entropy of entanglement of the state to be distilled is shown in red. The dotted lines interpolate linearly between neighbouring points. The value of the depolarising probability on the $x$-axis is given by Eq.~\eqref{DepolarisingProbability}.}
\label{fig:EffDepthHalf}
\end{figure}
\begin{figure}[ht]
\centering
\vspace{0.0cm}
\includegraphics[width=0.48\textwidth]{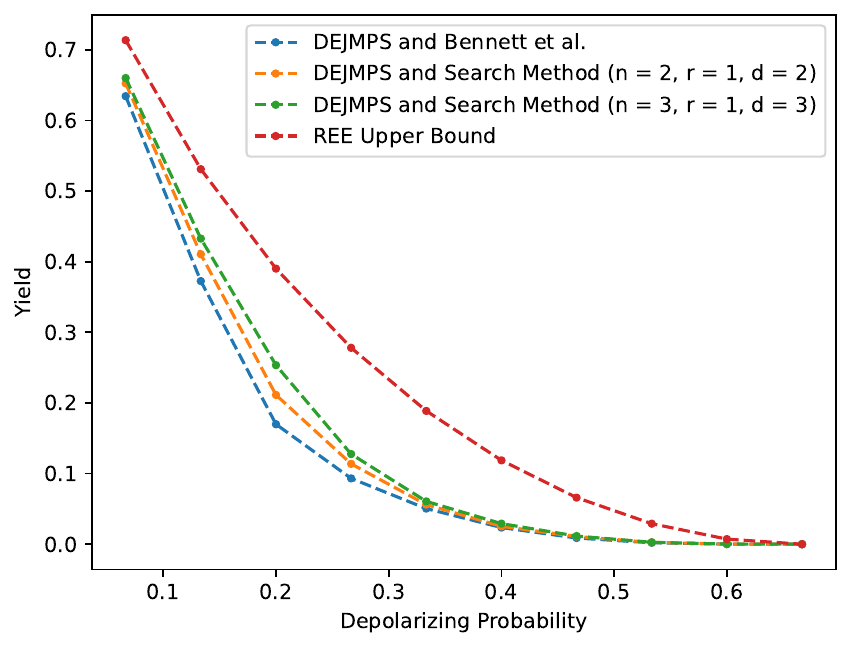}
\caption{The average yield of our search-based protocol with $\left(n, r, d\right) \in \left\{\left(2, 1, 2\right), \left(3, 1, 3\right)\right\}$ when applied to Werner states with Bell fidelity in $\left\{0.5, 0.55, 0.6, 0.65, \dots, 0.95\right\}$. The performance of the asymptotic protocol of Ref.~\cite{BBPSSW}, which is equivalent to our protocol with $\left(n, r\right) = \left(1, 1\right)$, is also shown. All asymptotic protocols are preceded by an optimal number of iterations of the protocol of Ref.~\cite{DEJMPS}. An upper bound, the relative entropy of entanglement of the state to be distilled is shown in red. The dotted lines interpolate linearly between neighbouring points. The value of the depolarising probability on the $x$-axis is given by Eq.~\eqref{DepolarisingProbability}.}
\label{fig:EffDepthOne}
\end{figure}

In Figs.~\ref{fig:EffDepthHalf} and~\ref{fig:EffDepthOne}, we fix $d_{\text{effective}}$ to be $\frac{1}{2}$ and $1$ respectively, varying $n$.
We then see that, across all Bell fidelities tested, our average yield increases as we increase $n$, albeit with only two values of $n$ tested per value of $d_{\text{effective}}$.
This is as we would expect based on the above discussion.

\par
\begin{figure}[h!]
\centering
\vspace{0.0cm}
\includegraphics[width=0.48\textwidth]{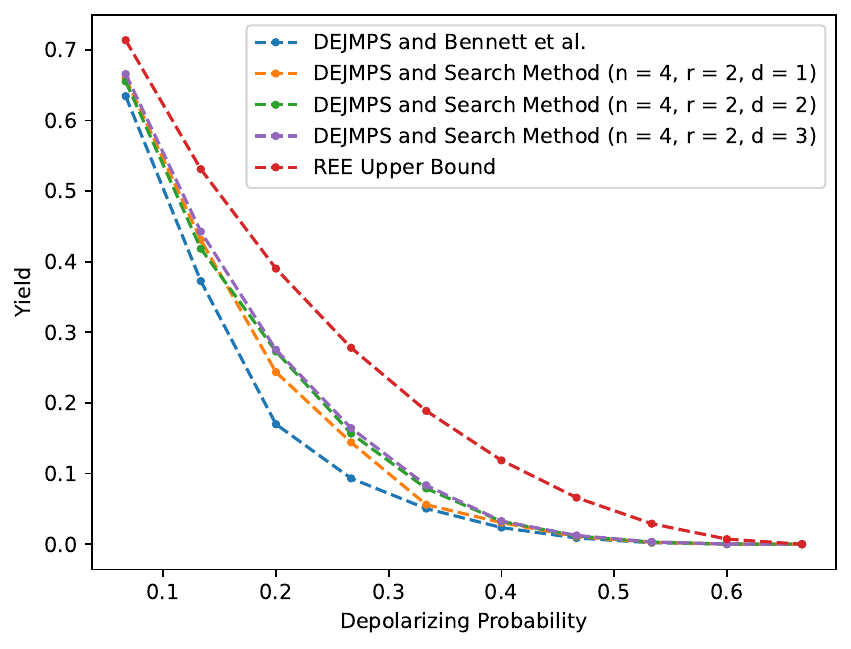}
\caption{The average yield of our search-based protocol with $\left(n, r\right) = \left(4, 2\right)$ and $d \in \left\{1, 2, 3\right\}$ when applied to Werner states with Bell fidelity in $\left\{0.5, 0.55, 0.6, 0.65, \dots, 0.95\right\}$. The performance of the asymptotic protocol of Ref.~\cite{BBPSSW} is also shown. All asymptotic protocols are preceded by an optimal number of iterations of the protocol of Ref.~\cite{DEJMPS}. An upper bound, the relative entropy of entanglement of the state to be distilled is shown in red. The dotted lines interpolate linearly between neighbouring points. The value of the depolarising probability on the $x$-axis is given by Eq.~\eqref{DepolarisingProbability}.}
\label{fig:VaryingDepth}
\end{figure}

Fig.~\ref{fig:VaryingDepth} shows the effect of varying $d$ on our protocol's average yield with fixed $n$ and $r$.
As expected, we generally observe better performance with larger $d$.
However, we also see that this is not guaranteed, as we sometimes achieve a higher yield with $d = 1$ than we do with $d = 2$.

\par
\begin{figure}[h]
\vspace{0.0cm}
\centering
\includegraphics[width=0.48\textwidth]{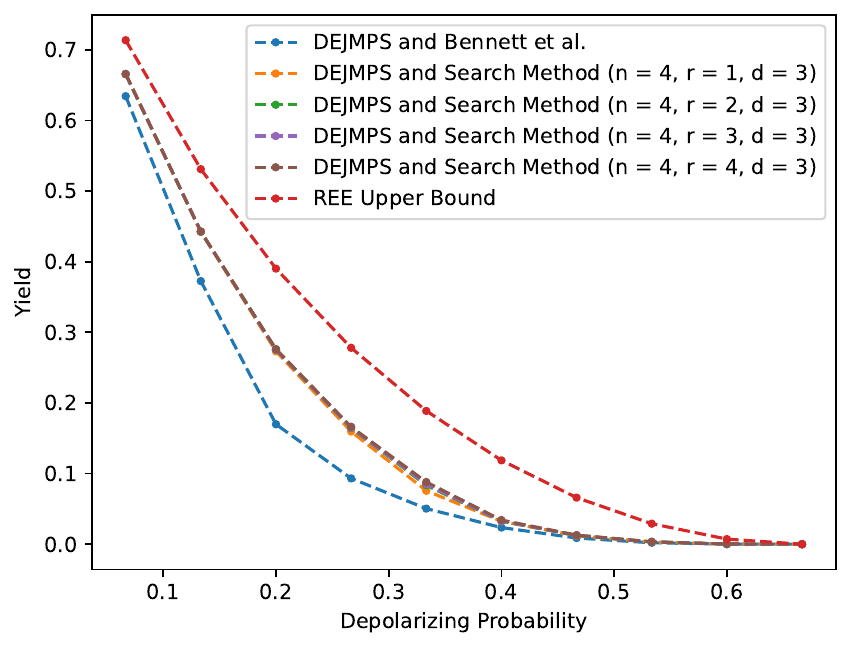}
\caption{The average yield of our search-based protocol with $\left(n, d\right) = \left(4, 3\right)$ and $r \in \left\{1, 2, 3, 4\right\}$ when applied to Werner states with Bell fidelity in $\left\{0.5, 0.55, 0.6, 0.65, \dots, 0.95\right\}$. The performance of the asymptotic protocol of Ref.~\cite{BBPSSW} is also shown. All asymptotic protocols are preceded by an optimal number of iterations of the protocol of Ref.~\cite{DEJMPS}. An upper bound, the relative entropy of entanglement of the state to be distilled is shown in red. The dotted lines interpolate linearly between neighbouring points. The value of the depolarising probability on the $x$-axis is given by Eq.~\eqref{DepolarisingProbability}.}
\label{fig:VaryingR}
\end{figure}

Fig.~\ref{fig:VaryingR} shows the effect of varying $r$ with $\left(n, d\right) = \left(4, 3\right)$.
Of our three parameters, $r$ seems to be the least significant when it comes to determining the average yield.
Increasing $r$ seems to improve our yield but not by much.
On the states with Bell fidelity $0.9$ and $0.95$, the value of $r$ made no difference to our average yield.
Indeed, if our states have little noise or $n$ is large, we would not expect to ever be left with just one qubit pair remaining as we applied our protocol and so $r$ should be unimportant in these circumstances.
\par
If $r = 1$, then it will take at most $2n$ parity checks to distil our system completely.
After all, with $n$ qubit pairs, there are $4^{n}$ possibilities for the state of our system and each AEM we perform rules out half of the remaining possibilities, while each BPM rules out three quarters of them.
Therefore, the best performance that our protocol could achieve would come from large $n$ with $d = 2n$.
In this case, $r$ would not matter, provided $n$ was sufficiently large for the given initial state.
Unfortunately, even with $r = 1$, the time complexity of Algorithm~\ref{alg:distillation_search} grows exponentially with respect to $nd$, as shown in Section~\ref{Complexity}.
In practice, then, we can improve our performance by setting $r > 1$ and using the recurrence protocol of Ref.~\cite{DEJMPS} when our states are noisy.

\subsection{Computational Complexity}\label{Complexity}

We will now consider the computational complexity of our search in the simplest case where $r = 1$.
We will assume that the distribution $\sigma$ is stored as a dense array of $4^m$ probabilities, that arithmetic on these probabilities takes constant time and that recursion is depth first, without memoisation.
We will begin with a worst-case theoretical evaluation of the time complexity before showing our empirical runtime scaling.
All $\mathcal{O}$ expressions are to be understood in the limit where $n$ and $d$ go to infinity.
\par
At a node with AEM and BPM shortlist dimensions $m_A$ and $m_B$, the number of candidate parity checks is $2^{m_A} + 2^{m_B} - 2$ and each check has two possible outcomes.
Initially, $m_A = m_B = 2n$, so there are $2(4^n - 1)$ candidate checks.
After each check, at least half of the shortlisted AEMs and BPMs are removed so, after $t$ parity checks, for some $t \in \mathbb{N}$, the number of remaining shortlisted parity checks is at most
\begin{equation}
  2\left(2^{2n-t} - 1\right).
\end{equation}
\par
Let $D = \min\left(\left\{d, 2m\right\}\right)$. The number of candidate--outcome paths explored by one search is at most
\begin{equation}
  P_{m}\left(d\right) = \prod_{j=0}^{D-1} 4\left(2^{2m-j} - 1\right) \leq 2^{\frac{D\left(4m+5-D\right)}{2}}\text{.}
\end{equation}
Each node costs $\mathcal{O}(m 4^{m})$ per candidate to compute outcome probabilities and conditional distributions, and $\mathcal{O}(m^{3})$ per candidate to update the shortlists.
One search therefore takes
\begin{equation}
  \mathcal{O}\left(\left(m4^{m}\right)P_{m}\left(d\right) \right)
\end{equation}
time.
The time complexity of the full evaluation based on Eqs. \eqref{AEMRecursion} and \eqref{BPMRecursion} is of the same order as the first search and is therefore given by
\begin{equation}
  \mathcal{O}\left(n2^{\frac{4nD+5D+4n-D^{2}}{2}}\right)\text{,}\label{FullComplexity}
\end{equation}
where $D = \min\left(\left\{d, 2n\right\}\right)$.
\par
\begin{figure}
    \vspace{-0.95cm}
    \centering
    \includegraphics[width=0.48\textwidth]{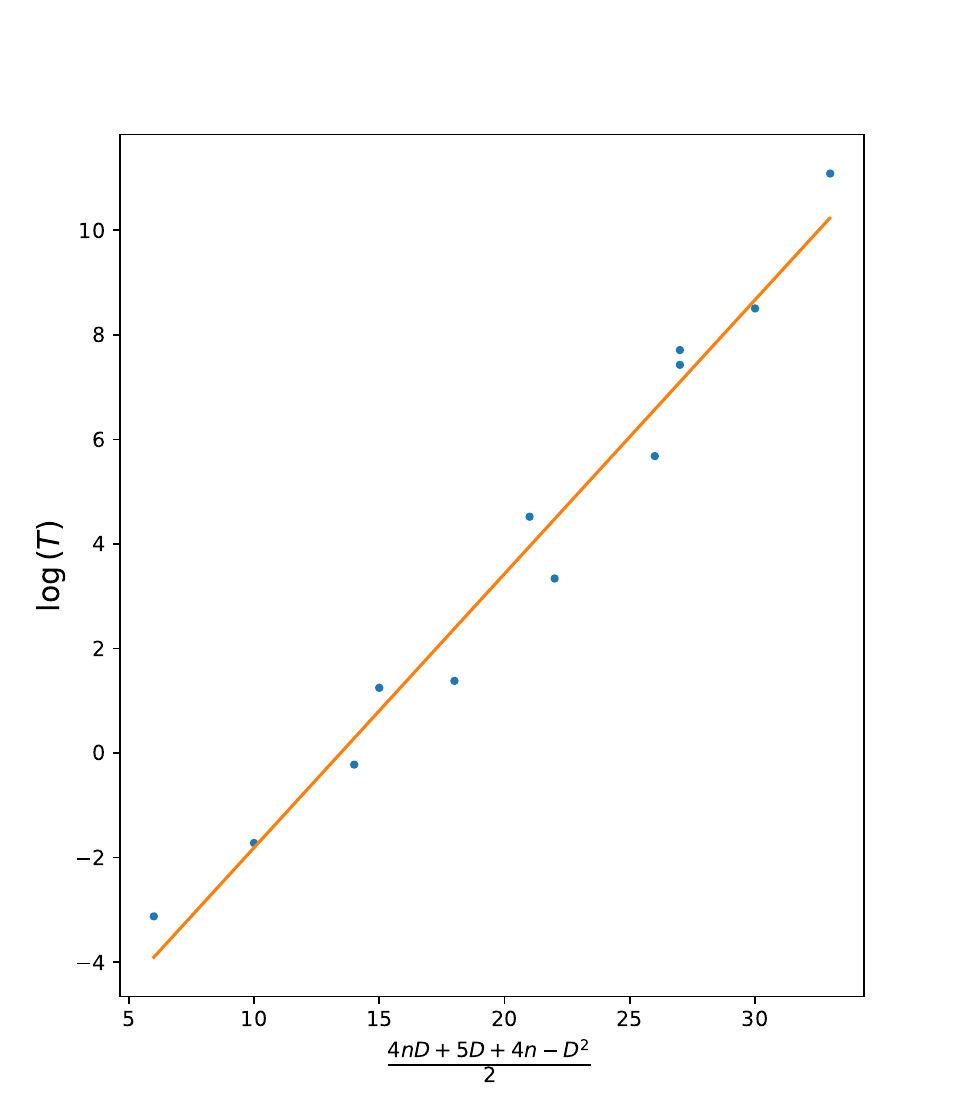}
    \vspace{-0.4cm}
    \caption{The natural logarithm of $T$, the runtime in seconds to calculate our yields for Figs. \ref{fig:Depolarising} to \ref{fig:VaryingR}, against the exponent in Eq.~\eqref{FullComplexity}. The data shown in blue all had $r = 1$. With $d = 1$, we used $n \in \left\{1, 2, \dots, 7\right\}$, with $d = 2$, $n \in \left\{2, 3, 4, 5\right\}$ and, with $d = 3$, $n = 3$. The orange line has a gradient of $0.524$ and a $y$-intercept of $-7.05$. The coefficient of determination between the empirical data and this line of best fit was $\simeq 0.973$.}
    \label{fig:runtime}
\end{figure}

In practice, our runtimes did not increase as fast as the worst-case theoretical analysis suggests.
Nevertheless, we saw similar exponential behaviour, as shown in Fig.~\ref{fig:runtime}.
Our line of best fit's gradient of $0.524$ corresponds to scaling roughly on the order of $1.69^{\frac{4nD+5D+4n-D^{2}}{2}}$.
However, this may depend on the initial state.
It is not surprising that real runtimes scale less quickly than Eq.\eqref{FullComplexity} would suggest.
For one thing, when we apply a BPM, typically only a quarter of candidate parity checks will remain eligible, rather than the half assumed in the worst-case theoretical analysis.

\section{Conclusion}

We have described a technique for performing entanglement distillation on qubit pairs.
From our empirical results, this technique appears to be significantly more effective than previous approaches when applied to Werner states with a wide range of Bell fidelities.
Our new protocols thereby allow us to improve upon the best previously-discovered lower bounds for the two-way-assisted quantum capacities of qubit depolarising channels across a wide range of depolarising probabilities.
Although not shown here, similar improvements can be seen for general qubit Pauli channels with a variety of parameters.
\par
This improved performance was achieved despite the fact that our search was restricted to parity check vectors significantly shorter than those that are used in the protocol of Ref.~\cite{HDDM}.
It may be that, if $d$ is taken to $\infty$ with $d = 2n$, the yield of the protocol will converge to the distillable entanglement of the initial state for all Bell-diagonal states.
Unfortunately, the size of the search space grows exponentially with the protocol's parameters, making it difficult to perform a much broader and deeper search.
\par
Because twirling can destroy entanglement, we do not expect this technique to be effective on states that are not approximately Bell-diagonal.
For instance, our technique did not provide similar improvements over previously-established lower bounds on the two-way-assisted quantum capacities of amplitude damping channels.
On states which are almost pure ebits, we expect our protocol will only be effective with large $n$, which may make the search computationally infeasible.

\smallskip
\noindent \textbf{Acknowledgements.~}
The authors acknowledge support from UKRI via the Integrated Quantum Networks Research Hub (IQN, EP/Z533208/1). 

\FloatBarrier

\newpage
\appendix
\section{Algorithmic Description}\label{Psuedocode}

There is some ambiguity in the protocol described in Sec.~\ref{RecedingHorizonSearch}.
In particular, there may be times when multiple candidate parity checks tie for the best $D$ or $E$ value.
For instance, when we search to a depth of $1$, all of the $D$ values will be the same.
The results of Sec.~\ref{Applications} were produced using the EvaluateCost procedure of Algorithm~\ref{alg:distillation_search}. 
In the event of a draw, this procedure favours BPMs over AEMs, favours BPMs that reduce the entropy of our probability distribution over equivalence classes as much as possible and favours AEMs whose measurement outcomes are as uncertain as possible. Whenever we refer to an ordering on lists in Algorithm~\ref{alg:distillation_search}, the ordering is lexicographic.

\begin{widetext}

\noindent\rule{\textwidth}{0.8pt}

\algorithmcaption{Distillation Protocol Search with Receding Horizon}
\label{alg:distillation_search}

\noindent\rule{\textwidth}{0.4pt}

\begin{algorithmic}[1]

\Require State $\sigma$,
AEM basis $\tilde{A}$,
BPM basis $\tilde{B}$,
lookahead depth $d$,
block size $r$,
boolean flag \textit{SearchMode},
list $\mathcal{R}$ of single-pair states that started enclosing blocks,
list $\mathcal{G}$ of corresponding guesses.
The first call uses empty lists $\mathcal{R}=\mathcal{G}=()$.

\Ensure Expected measurement cost
$C(\sigma,\tilde{A},\tilde{B})$
to distil ebits.

\vspace{0.2cm}

\Procedure{GenerateCandidates}
{$\tilde{A},\tilde{B}$}

    \State
    $\mathbb{C} \gets \text{Empty List}$

    \State \textbf{Generate AEMs:}

    \State
    $m_A \gets \dim(\tilde{A})$

    \For{$k=1$ \textbf{to} $2^{m_A}-1$}

        \State
        $x \gets
        \text{binary representation of } k
        \text{ padded to length }m_A$

        \State
        $c \gets
        \sum_{i=1}^{m_A}
        x_i\tilde{A}_i
        \pmod 2$

        \State
        Append $c$ to $\mathbb{C}$

    \EndFor

    \State \textbf{Generate BPMs:}

    \State
    $m_B \gets \dim(\tilde{B})$

    \For{$k=1$ \textbf{to} $2^{m_B}-1$}

        \State
        $x \gets
        \text{binary representation of } k
        \text{ padded to length }m_B$

        \State
        $c \gets
        \sum_{i=1}^{m_B}
        x_i\tilde{B}_i
        \pmod 2$

        \State
        Append $c$ to $\mathbb{C}$

    \EndFor

    \State
    \Return $\mathbb{C}$

\EndProcedure

\vspace{0.2cm}

\Procedure{EvaluateCost}
{$\sigma$,
 $\tilde{A}$,
 $\tilde{B}$,
 $d$,
 $r$,
 \textit{SearchMode},
 $\mathcal{R}$,
 $\mathcal{G}$}

    \State
    $m \gets
    \text{current number of surviving pairs}$

    \State
    $H^{\star}(\sigma)
    \gets
    \min(m,H(\sigma))$
    \Comment{Bounded entropy heuristic}

    \If{$H^{\star}(\sigma)=0$}
        \State \Return $0$
        \Comment{State is pure; distillation complete}
    \EndIf

    \If{\textit{SearchMode} \textbf{and} $d=0$}

        \State
        \Return $H^{\star}(\sigma)$
        \Comment{Halt lookahead and return heuristic}

    \EndIf

    \State
    \textbf{Phase 1: Block Formation, Cycle Detection \& Algebraic Extraction}

    \If{
        \textit{SearchMode}
        \textbf{and}
        $m=1$
        \textbf{and}
        $r>1$
    }

        \State
        \Return
        $\frac{1}{r}
        \times
        \Call{EvaluateCost}
        {
            \sigma^{\otimes r},
            \tilde{K}_r,
            \tilde{K}_r,
            d,
            r,
            \text{True},
            \mathcal{R},
            \mathcal{G}
        }$
        \Comment{Hypothetical block; continue searching}

    \EndIf

    \If{$m=1$}

        \For{$\ell=1$ \textbf{to} $|\mathcal{R}|$}

            \If{$\sigma\cong\mathcal{R}[\ell]$}

                \State
                \Return $\mathcal{G}[\ell]$
                \Comment{Returned to an earlier state}

            \EndIf

        \EndFor

    \EndIf

    \If{
        $m=1$
        \textbf{and}
        $r>1$
    }

        \State
        $\sigma_r
        \gets
        \sigma^{\otimes r}$
        \Comment{Form new block of size $r$}

        \State
        $C_0
        \gets
        \frac{1}{r}
        \times
        \Call{EvaluateCost}
        {
            \sigma_r,
            \tilde{K}_r,
            \tilde{K}_r,
            d,
            r,
            \text{False},
            \mathcal{R}\frown(\sigma),
            \mathcal{G}\frown(0)
        }$

        \State
        $C_1
        \gets
        \frac{1}{r}
        \times
        \Call{EvaluateCost}
        {
            \sigma_r,
            \tilde{K}_r,
            \tilde{K}_r,
            d,
            r,
            \text{False},
            \mathcal{R}\frown(\sigma),
            \mathcal{G}\frown(1)
        }$

        \State
        $\beta
        \gets
        C_1-C_0$
        \Comment{Copies of $\sigma$ returned per copy used}

        \State
        \Return
        $C_0/(1-\beta)$
        \Comment{Sum of geometric series}

    \EndIf

    \State
    \textbf{Phase 2: Candidate Generation \& Heuristic Sorting}

    \State
    $\mathbb{C}
    \gets
    \Call{GenerateCandidates}
    {\tilde{A},\tilde{B}}$

    \For{$j=1$ \textbf{to} $|\mathbb{C}|$}

        \State
        $c \gets \mathbb{C}[j]$

        \State
        $R_c \gets w(c)$
        \Comment{Check reduction for AEMs}

        \State
        $H_c
        \gets
        \sum_{i\in\{0,1\}}
        \Pr(\sigma_c=i)
        H(\sigma_i)$
        \Comment{Average remaining Shannon entropy}

        \State
        $\tau_c
        \gets
        w(c)
        +
        \sum_{i\in\{0,1\}}
        \Pr(\sigma_c=i)
        H^*(\sigma_i)$
        \Comment{One-step heuristic}

        \State
        $\mathbf{S}_c
        \gets
        (\tau_c,j)$
        \Comment{Sorting key to ensure stability}

    \EndFor

    \State
    Sort $\mathbb{C}$ in ascending order by
    $\mathbf{S}_c$
    to form the ordered list $L$.

    \State
    \textbf{Phase 3: Pruned Deep Search (Hypothetical Lookahead)}

    \State
    $\mathbf{V}_{\text{best}}
    \gets
    (\infty,\infty,\infty,\infty)$

    \State
    $c_{\text{best}}
    \gets
    \emptyset$

    \For{$k=1$ \textbf{to} $|L|$}

        \State
        $c \gets L[k]$

        \State
        Obtain updated bases
        $\tilde{A}',\tilde{B}'$
        after applying $c$

        \State
        $E_c \gets w(c)$

        \For{$i\in\{0,1\}$}

            \If{$\Pr(\sigma_c=i)>0$}

                \State
                $E_c
                \gets
                E_c
                +
                \Pr(\sigma_c=i)
                \times
                \Call{EvaluateCost}
                {
                    \sigma_i,
                    \tilde{A}',
                    \tilde{B}',
                    d-1,
                    r,
                    \text{True},
                    \mathcal{R},
                    \mathcal{G}
                }$

            \EndIf

            \If{$E_c>\mathbf{V}_{\text{best}}[0]$}

                \State \textbf{break}
                \Comment{Branch \& Bound pruning}

            \EndIf

        \EndFor

        \If{$c$ is a BPM}

            \State
            $t_c\gets0$

            \State
            $\nu_c\gets H_c$

        \Else

            \State
            $t_c\gets1$

            \State
            $\nu_c\gets-R_c$

        \EndIf

        \State
        $\mathbf{V}_c
        \gets
        (E_c,t_c,\nu_c,-k)$
        \Comment{Construct lexicographical vector}

        \If{
            $\mathbf{V}_c
            <
            \mathbf{V}_{\text{best}}$
        }

            \State
            $\mathbf{V}_{\text{best}}
            \gets
            \mathbf{V}_c$

            \State
            $c_{\text{best}}
            \gets
            c$

        \EndIf

    \EndFor

    \State
    \textbf{Phase 4: Physical Traversal (Execution)}

    \If{\textit{SearchMode} \textbf{is True}}

        \State
        \Return
        $\mathbf{V}_{\text{best}}[0]$
        \Comment{Pass estimate up the hypothetical tree}

    \Else

        \State
        Obtain
        $\tilde{A}'$,
        $\tilde{B}'$,
        $\Pr(\sigma_{c_{\text{best}}}=i)$,
        and $\sigma_i$
        for the chosen $c_{\text{best}}$

        \State
        \Return
        $w(c_{\text{best}})$

        \Statex
        \qquad\qquad
        $+
        \sum_{i\in\{0,1\}}
        \Pr(\sigma_{c_{\text{best}}}=i)
        \times
        \Call{EvaluateCost}
        {
            \sigma_i,
            \tilde{A}',
            \tilde{B}',
            d,
            r,
            \text{False},
            \mathcal{R},
            \mathcal{G}
        }$

    \EndIf

\EndProcedure

\end{algorithmic}
\vspace{-0.2cm}

\noindent\rule{\textwidth}{0.8pt}

\end{widetext}


\section{Modification to Eq.~(13) of Asymptotic Adaptive Bipartite Entanglement-Distillation Protocol}

In order to reproduce the results of Ref.~\cite{HDDM}, we had to make a slight modification to Eq.~(13) of the same paper.
In most cases, this modification makes no difference.
However, when, to use the notation of Ref.~\cite{HDDM}, $u\left(A\right) = 0$, the two formulae – our modified version and the original – are not equivalent.
Throughout this appendix, we will adopt the notation and terminology of Ref.~\cite{HDDM}.
\par
At each stage, the protocol of Sec.~VD of Ref.~\cite{HDDM} decomposes a large number of $0^{\left(2m\right)}$ and $1^{\left(2m\right)}$ states into $0^{\left(m\right)}$ and $1^{\left(m\right)}$ states via a series of rounds of BPMs.
Every $0^{\left(2m\right)}$ state is ordered as described in Ref.~\cite{HDDM} and, for every $i \in \mathbb{N}$, it is only possible for a BPM to be performed on half of a $0^{\left(2m\right)}$ state during the $i$th round if the state lies in the first $\frac{1}{2^{i}}$ of all $0^{\left(2m\right)}$ states.
If, for any $i \in \mathbb{N}$, a $0^{\left(2m\right)}$ does lie in the first $\frac{1}{2^{i}}$ of all $0^{\left(2m\right)}$ states, the probability of a BPM being performed on half of it is $z_{i}$.
Now, we define
\begin{equation}
    u'\left(A\right) = \lfloor -\log_{2}\left(U\left(A\right)\right)\rfloor\text{.}
\end{equation}
For every $i \in \left\{1, 2, \dots, u'\left(A\right)\right\}$, every state in $A$ lies in the first $\frac{1}{2^{i}}$ of all $0^{\left(2m\right)}$ states.
For every $i \in \left\{u'\left(A\right)+1, u'+2, \dots, l\left(A\right)\right\}$, the fraction of states in $A$ that lie in the first $\frac{1}{2^{i}}$ of all $0^{\left(2m\right)}$ states is $\frac{2^{-i}-L\left(A\right)}{U\left(A\right)-L\left(A\right)}$\text{.}
Finally, for every $i \in \left\{l\left(A\right)+1, l\left(A\right)+2, \dots\right\}$, there are no states in $A$ that lie in the first $\frac{1}{2^{i}}$ of all $0^{\left(2m\right)}$ states.
Therefore, $\eta\left(0^{\left(2m\right)} \mid A\right)$ should be
\begin{equation}
     \sum_{i=1}^{u'\left(A\right)}z_{i} + \sum_{i = u'\left(A\right) + 1}^{l\left(A\right)}\frac{2^{-i}-L\left(A\right)}{U\left(A\right)-L\left(A\right)}z_{i}\text{.}\label{etaFormula}
\end{equation}
\par
Now, provided $\log_{2}\left(U\left(A\right)\right)$ is not an integer, we have
\begin{equation}
    u'\left(A\right) = u\left(A\right) - 1
\end{equation}
so that Eq.~\eqref{etaFormula} is equivalent to the right-hand side of Eq.~(13) of Ref.~\cite{HDDM}.
Moreover, if we have $-\log_{2}\left(U\left(A\right)\right) \in \mathbb{N}$, then the right-hand side of Eq.~(13) of Ref.~\cite{HDDM} is
\begin{align}
    &\sum_{i=1}^{u\left(A\right)-1}z_{i} + \sum_{i = u\left(A\right)}^{l\left(A\right)}\frac{2^{-i}-L\left(A\right)}{U\left(A\right)-L\left(A\right)}z_{i} \nonumber \\
    = &\sum_{i=1}^{u'\left(A\right)-1}z_{i} + \sum_{i = u'\left(A\right)}^{l\left(A\right)}\frac{2^{-i}-L\left(A\right)}{U\left(A\right)-L\left(A\right)}z_{i}  \nonumber \\
    = &\sum_{i=1}^{u'\left(A\right)}z_{i} + \sum_{i = u'\left(A\right) + 1}^{l\left(A\right)}\frac{2^{-i}-L\left(A\right)}{U\left(A\right)-L\left(A\right)}z_{i}\text{,}
\end{align}
again in accordance with Eq.~\eqref{etaFormula}.
However, when $-\log_{2}\left(U\left(A\right)\right) = 0$ or, equivalently, $U\left(A\right) = 1$, there is a discrepancy between the two formulae.
In that case, the right-hand side of Eq.~(13) of Ref.~\cite{HDDM} is
\begin{equation}
    \sum_{i=1}^{-1}z_{i} + z_{0} + \sum_{i = 1}^{l\left(A\right)}\frac{2^{-i}-L\left(A\right)}{1-L\left(A\right)}z_{i}\text{,}
\end{equation}
while Eq.~\eqref{etaFormula} is
\begin{equation}
    \sum_{i = 1}^{l\left(A\right)}\frac{2^{-i}-L\left(A\right)}{1-L\left(A\right)}z_{i}\text{.}
\end{equation}
If we interpret $\sum_{i=1}^{-1}z_{i}$ as $-z_{0}$, terms with indices lower than the lower index of the summation being multiplied by $-1$, then the two formulae are once again in agreement.
However, we would have thought the standard evaluation of $\sum_{i=1}^{-1}z_{i}$ would be $0$, in which case Eq.~\eqref{etaFormula} and the right-hand side of Eq.~(13) of Ref.~\cite{HDDM} are not equivalent, as $z_{0}$ is non-zero in general.


\bibliography{references}

\end{document}